\documentclass[aps,pra,reprint,superscriptaddress,showpacs]{revtex4-1}

\usepackage{latexsym}
\usepackage{graphics}
\usepackage{amsmath}
\usepackage{color}
\usepackage{graphicx}

\newcommand {\bfr} {{\bf r}}
\newcommand {\bfE} {{\bf E}}

\renewcommand {\d} {{\rm d}}

\newcommand {\rmC} {{\rm C}}

\newcommand {\om} {\omega}

\newcommand {\s} {{\sigma}}

\newcommand {\calA} {{\cal A}}

\newcommand {\calN} {{\cal N}}

\begin{document}

\title{Photoionization of multishell fullerenes studied by ab initio and model approaches [Eur. Phys. J. D 70 (2016) 221]}

\author{Alexey Verkhovtsev}
\email[]{verkhovtsev@iff.csic.es}
\altaffiliation{On leave from A.F. Ioffe Physical-Technical Institute of Russian Academy of Sciences, 
Politekhnicheskaya 26, 194021 St. Petersburg, Russia}
\affiliation{MBN Research Center, Altenh\"oferallee 3, 60438 Frankfurt am Main, Germany}
\affiliation{Instituto de F\'{\i}sica Fundamental, CSIC, Serrano 113-bis, 28006 Madrid, Spain}

\author{Andrei V. Korol}
\affiliation{MBN Research Center, Altenh\"oferallee 3, 60438 Frankfurt am Main, Germany}
\affiliation{St. Petersburg State Maritime Technical University, Leninskii ave. 101, 198262 St. Petersburg, Russia}

\author{Andrey V. Solov'yov}
\altaffiliation{On leave from A.F. Ioffe Physical-Technical Institute of Russian Academy of Sciences,
Politekhnicheskaya 26, 194021 St. Petersburg, Russia}
\affiliation{MBN Research Center, Altenh\"oferallee 3, 60438 Frankfurt am Main, Germany}


\begin{abstract}
Photoionization of two buckyonions, C$_{60}$@C$_{240}$ and C$_{20}$@C$_{60}$, is investigated
by means of time-dependent density-functional theory (TDDFT).
The TDDFT-based photoabsorption spectrum of C$_{60}$@C$_{240}$, calculated in a broad photon
energy range, resembles the sum of spectra of the two isolated fullerenes, thus illustrating
the absence of strong plasmonic coupling between the fullerenes which was proposed earlier.
The calculated spectrum of the smaller buckyonion, C$_{20}$@C$_{60}$, differs significantly
from the sum of the cross sections of the individual fullerenes because of strong geometrical
distortion of the system.
The contribution of collective electron excitations arising in individual fullerenes
is evaluated by means of plasmon resonance approximation (PRA).
An extension of the PRA formalism is presented, which allows for the study of collective
electron excitations in multishell fullerenes under photon impact.
An advanced analysis of photoionization of buckyonions, performed using modern computational
and analytical approaches, provides valuable information on the response of complex molecular
systems to the external electromagnetic field.
\end{abstract}

\maketitle

\section{Introduction}

Formation and dynamics of electron excitations in fullere\-nes, their derivatives
and other carbon-based nanoscale systems like polycyclic aromatic hydrocarbons (PAHs)
ha\-ve been widely studied, both experimentally and theoretically, during the past decades
\cite{Wopperer_2015_PhysRep.562.1, Lepine_2012_JPB.48.122002, Solovyov_2005_IJMPB.19.4143, Verkhovtsev_2012_EPJD.66.253,Kilcoyne_2010_PRL.105.213001,Tribedi_2015_PRA.92.060701R}.
A particular attention has been paid to ionization of a C$_{60}$ fullerene under the
photon, electron, and ion impact
\cite{Hertel_1992_PRL.68.784, Berkowitz_1999_JCP.111.1446, Scully_2005_PRL.94.065503, Mikoushkin_1998_PRL.81.2707, Bolognesi_2012_EPJD.66.254, Schueler_2015_PRA.92.021403R, Verkhovtsev_2015_EPJD.69.116, Baral_2016_PRA.93.033401}.
Because of their high symmetry and stability, these molecules have been of significant
fundamental interest aimed at better understanding the photon- and charged-particle-induced
processes in complex many-particle systems.
The understanding of the mechanisms of electron emission from nanoscale systems exposed
to ionizing radiation is a key issue in a wide range of physical and chemical
processes~\cite{Brun_2009_JPCB.113.10008,Verkhovtsev_2015_PRL.114.063401}.

Although the structure and dynamics of pristine fullere\-nes have been widely explored,
much less attention has been paid to more complex systems, namely multishell fullerenes
or buckyonions -- concentric carbon nanostructures composed of several
nested molecules~\cite{Ugarte_1992_Nature.359.707,Ugarte_1995_Carbon.33.989}.
Unlike fulle\-renes and carbon nanotubes, the properties of carbon buckyonions are still
not well understood.
Accurate theoretical studies of the structure and dynamical properties of these systems
are also rather limited because of the large number of constituent atoms and related high
computational costs.

Several papers have been devoted to the study of the static dipole polarizability of carbon
buckyonions~\cite{Iglesias_2003_JCP.118.7103,Zope_2008_JPB.41.085101}.
Density-functional theory (DFT) calculations~\cite{Zope_2008_JPB.41.085101} revealed that the
static dipole polarizability of the C$_{60}$@C$_{240}$ onion is very similar to that of the
isolated C$_{240}$, so that the inner fullerene is almost completely screened by the outer shell.
References~\cite{Dolmatov_2008_PRA.78.013415,Amusia_2009_PRA.80.032503}
were devoted to a theoretical study of photoionization and photoelectron angular distribution
asymmetry parameters of an atom $A$ confined in spherical multiwalled fullerenes,
$A$@C$_{60}$@C$_{240}$ and $A$@C$_{60}$@C$_{240}$@C$_{540}$.
In these works, the electronic structure of a confined atom was calculated explicitly, e.g.,
within the Hartree-Fock approximation, while the fullerene shells were modeled by a set of
spherical zero- or finite-width attractive potentials.

Finally, only a few works have investigated photoabsorption spectra of buckyonions
\cite{Ruiz_2003_JCP.120.6163,McCune_2011_JPB.44.241002,Casella_2013_PCCP.15.18030}.
Similar to pristine fullerenes, nanotubes and PAHs,
photoabsorption and electron energy loss spectra of buckyonions are characterized
by prominent plasmon resonances formed due to collective excitations of delocalized
$\sigma$ and $\pi$ electrons~\cite{Cabioch_1997_EPL.38.471,Chhowalla_2003_PRL.90.155504}.
In Ref.~\cite{Ruiz_2003_JCP.120.6163}, Ruiz {\it et al.} presented a theoretical model for the calculation
of the photoabsorption spectra of spherical $N$-shell carbon buckyonions in the low photon
energy region (below 10~eV) dominated by the $\pi$-plasmon.
The effect due to the $\pi$-plasmon was evaluated in this model based on the electronic
structure of the system provided by the H\"uckel single-electron model.
In Ref.~\cite{McCune_2011_JPB.44.241002}, the photoionization of a bilayer C$_{60}$@C$_{240}$
onion was studied by means of the time-dependent local-density approximation, while the
jellium model, that is, a uniform smearing of the valence electron density over a finite-width
spherical shell, was used to represent the electronic structure of each fullerene.
The calculated photoionization spectrum of C$_{60}$@C$_{240}$ showed a significant
redistribution of the oscillator strength density and the emergence of two new resonances,
as compared to the sum of the cross sections of the pristine systems.
These effects were explained in terms of a strong inter-fullerene coupling leading to
hybridization of the electronic states of individual fullerenes and the formation of four
cross-over plasmons~\cite{McCune_2011_JPB.44.241002}.
In a recent work~\cite{Casella_2013_PCCP.15.18030}, the photoabsorption spectra of
C$_{60}$@C$_{240}$ and C$_{60}$@C$_{180}$ onions were calculated by means of time-dependent
DFT (TDDFT) in the visible-near UV region (up to 5 eV).
The resulting spectrum of C$_{60}$@C$_{240}$ was characterized by a simple overlap of the
spectra of isolated C$_{60}$ and C$_{240}$ because of a weak mutual perturbation of the
two fullerenes upon encapsulation.

In this paper, we evaluate the photoabsorption spectra of two buckyonions, C$_{60}$@C$_{240}$
and C$_{20}$@C$_{60}$, by means of the two complementary theoretical approaches.
The TDDFT method is used to calculate the spectra in a broad photon energy range up to 100~eV.
The results of TDDFT-based calculations are compared with those based on the plasmon resonance
approximation (PRA)~\cite{Solovyov_2005_IJMPB.19.4143,Verkhovtsev_2012_EPJD.66.253,Connerade_AS_PRA.66.013207}
in order to map the well-resolved features of the spectra to particular collective electron
excitations.
The PRA formalism is extended allowing for the study of collective electron excitations
in multishell fullerenes under photon impact.
We demonstrate that the spectrum of the C$_{60}$@C$_{240}$ buckyonion corresponds to the
sum of the spectra of the two isolated fullerenes, thus indicating the absence of strong
inter-fullerene coupling.
On the contrary, the photoabsorption spectrum of C$_{20}$@C$_{60}$ differs significantly from
the corresponding sum of the isolated molecules due to strong geometrical distortion of the system.
These results provide valuable information about a fundamental problem of the formation
and interplay of collective electron excitations in complex nanoscale systems.

The atomic system of units, $m_e = |e| = \hbar = 1$, is used throughout the paper unless
otherwise indicated.

\section{Theory and computational details}

\subsection{Time-dependent density-functional theory}

The TDDFT-based calculations of isolated fullerenes and buckyonions have been performed in the
linear regime within the dipole approximation~\cite{Walker_2006_PRL.96.113001,Rocca_2008_JCP.128.154105}.
Within this framework, the external potential $v_{\rm ext}(\bfr,t)$ acting on a
system is represented as a sum of a time-independent part, $v_{\rm ext}^0(\bfr)$,
and a time-dependent perturbation $v_{\rm ext}^{\prime}(\bfr, t)$.
The time evolution of the electron density, $\rho(\bfr, t)$, is then represented
as a sum of the unperturbed ground-state density, $\rho_0(\bfr)$, and the variation
$\delta\rho(\bfr, t)$, which arises due to $v^{\prime}_{\rm ext}(\bfr, t)$.

Performing the Fourier transform of time-depen\-dent quantities, one gets the
response of the system to an external perturbation in the frequency representation.
For the external perturbation $v_{\rm ext}^{\prime}(\bfr,\om) = - {\bf E}(\om) \cdot {\bfr}$
due to a uniform electric field, the Fourier transform of the induced dipole moment reads
as follows:
\begin{equation}
d_i(\om) = \sum_j \alpha_{ij}(\om) \, E_j(\om) \ ,
\end{equation}
where $i, j$ denote the Cartesian components, $\alpha_{ij}(\om)$ is the dynamical
polarizability tensor which describes the linear response of the dipole to the external
electric field:
\begin{equation}
\alpha_{ij}(\om) =
- \int r_i \, \chi(\bfr,\bfr^{\prime},\om) \, r_j^{\prime} \, {\rm d}\bfr \, {\rm d}\bfr^{\prime} \ ,
\end{equation}
$\chi(\bfr,\bfr^{\prime},\om)$ is the generalized frequency-dependent susceptibility
of the system, and $r_i$ and $r_j^{\prime}$ are the components of the position operators
$\bfr$ and $\bfr^{\prime}$.
The photoabsorption cross section is related to the imaginary part of $\alpha_{ij}(\om)$ through
\begin{equation}
\sigma(\om) = \frac{4\pi\om}{3c} \sum_j {\rm Im} \left[ \alpha_{jj}(\om) \right] \ ,
\end{equation}
where $c$ is the speed of light, and the summation is performed over
the diagonal elements of the polarizability tensor.

The performed calculations rely on the approach introduced in Refs.~\cite{Walker_2006_PRL.96.113001,Rocca_2008_JCP.128.154105}, which is based
on a superoperator formulation of TDDFT.
It allows for the calculation of the dynamic polarizability by means of an efficient
Lanczos method.
In this approach, the polarizability of a many-electron system is expressed as \cite{Malcioglu_2011_CompPhysCommun.182.1744}:
\begin{equation}
\alpha_{ij}(\om) = {\rm Tr} \left( \hat{X}_i \, \hat{\rho}_j^{\prime}(\om) \right) \ ,
\label{alpha_tensor}
\end{equation}
where the hat symbols indicate quantum mechanical operators, $\hat{X}_i$ is the $i$th
component of the position operator~$\hat{X}$, and
$\hat{\rho}_j^{\prime}(\om) = \hat{\rho}_j(\om) - \hat{\rho}_{0}$ is the response density matrix.
It is expressed via $\hat{\rho}_j(\om)$, that is the single-electron density matrix of the system
perturbed by an external homogeneous electric field polarized along the $j$th axis, and
$\hat{\rho}_{0}$, that is the density matrix describing the ground state.
The response density matrix can be expressed as the solution of the linearized quantum
Liouville equation~\cite{Walker_2006_PRL.96.113001,Rocca_2008_JCP.128.154105,Walker_2007_JCP.127.164106}:
\begin{equation}
(\om - \mathcal{L}) \cdot \hat{\rho}^{\prime}_j(\om) =
\left[ \hat{X}_j, \hat{\rho}_0 \right] \ ,
\label{Eq_qLiouville}
\end{equation}
where $\mathcal{L}$ is the so-called Liouvillian operator and the square brackets indicate a commutator.
The action of the Liouvillian $\mathcal{L}$ onto $\hat{\rho}^{\prime}(\om)$ is defined as
\begin{equation}
\mathcal{L} \cdot \hat{\rho}^{\prime}(\om) =
\left[ \hat{H}_0, \hat{\rho}^{\prime}(\om) \right] +
\left[ \hat{V}^{\prime}_{\rm Hxc}[\hat{\rho}^{\prime}], \hat{\rho}_0 \right] \ ,
\end{equation}
where $\hat{H}_0$ is the ground-state Kohn-Sham Hamiltonian calculated within the DFT approach
and $\hat{V}^{\prime}_{\rm Hxc}[\hat{\rho}]$ denotes the linear variation of the
electrostatic and exchange-corre\-lation potentials.
The coordinate representation of the latter operator is as follows:
\begin{equation}
v^{\prime}_{\rm Hxc}(\bfr,\om) =
\int \left( \frac{1}{|\bfr - \bfr^{\prime}|} + \kappa_{\rm xc}(\bfr, \bfr^{\prime}; \om) \right) \,
\rho^{\prime}(\bfr, \bfr^{\prime}; \om) {\rm d}\bfr^{\prime} \ ,
\end{equation}
and $\kappa_{\rm xc}$ is the so-called exchange-correlation kernel.
The polarizability tensor~(\ref{alpha_tensor}) is defined from the solution of Eq.~(\ref{Eq_qLiouville})
as the off-diagonal matrix element of the resolvent of the Liouvillian $\mathcal{L}$.
In the approach introduced in Refs.~\cite{Rocca_2008_JCP.128.154105,Malcioglu_2011_CompPhysCommun.182.1744,Walker_2007_JCP.127.164106},
this quantity is calculated using the Lanczos recursion method
(for details, see the above cited papers and references therein).

In this study, photoabsorption spectra of pristine and multishell fullerenes were
obtained for the systems with optimized geometries.
The optimization procedure was performed by means of Gaussian~09 package~\cite{g09}
utilizing the split-valence 6-31G(d) basis set and the local density approximation
(LDA)~\cite{PZ_1981_PRB.23.5048}.
The photoabsorption spectra of the optimized systems were obtained using the TDDFPT module
\cite{Malcioglu_2011_CompPhysCommun.182.1744} of the QuantumEspresso package \cite{Giannozzi_2009_JPCM.21.395502}.
The optimized structures were introduced into a supercell of $20 \times 20 \times 20~{\rm \AA}^3$.
Then, the system of Kohn-Sham equations was solved self-consistently for all valence electrons
(2$s^2$2$p^2$ electron in each carbon atom) of each system to calculate the ground-state
eigenvalues using a plane-wave approach~\cite{Giannozzi_2009_JPCM.21.395502}.
In the calculations, we used an ultrasoft pseudopotential~\cite{RRKJ_pseudopotential} which
substitutes real atomic orbitals in the core region with smooth nodeless pseudo-orbitals.
For the plane-wave calculations we used the kinetic energy cutoff of 30 Ry
for the wave functions and 180 Ry for the electron densities.

\subsection{Plasmon resonance approximation}
\label{PRA_pristine_fullerene}

The contribution of plasmon excitations to the photoabsorption spectra of isolated
fullerenes has been evaluated using the plasmon resonance approximation (PRA)
\cite{Solovyov_2005_IJMPB.19.4143, Verkhovtsev_2012_EPJD.66.253, Connerade_AS_PRA.66.013207}.
Within this approach, a single fullerene is represented as a spherically symmetric
system with a homogeneous charge distribution within the shell of a finite width,
$\Delta R = R_2 - R_1$, where $R_{1,2}$ are the inner and the outer radii of the molecule,
respectively~\cite{Lambin_Lukas_1992_PRB.46.1794, Oestling_1993_EPL.21.539, Lo_2007_JPB.40.3973, 
Verkhovtsev_2012_JPB.45.141002, Verkhovtsev_2013_JPCS.438.012011}.
In other words, a fullerene is modeled as a system with the electron density being constant
inside the spherical shell and equal to zero outside the shell.
The chosen value of the shell's width, $\Delta R = 1.5$~\AA, corresponds to the typical
size of the carbon atom \cite{Oestling_1993_EPL.21.539} and was successfully utilized in
our earlier studies, see, e.g., Refs.~\cite{Bolognesi_2012_EPJD.66.254, Verkhovtsev_2012_JPB.45.141002, 
Verkhovtsev_2013_PRA.88.043201}.
A question of validity of the rectangular electron density profile and its comparison with more
realistic densities in C$_{60}$ was addressed in a number of earlier studies, see, e.g.,
Refs.~\cite{Apell_1993_SolidStateCommun.87.219, Vasvari_1996_ZPhysB.100.223}.
In these works, it was found that the shape of the electron density distributions does not affect
the overall shape and the peak positions of the imaginary part of the dipole polarizability and
hence the photoionization cross section.

Within the PRA, the dynamical polarizability $\alpha(\om)$ has a resonance behavior in
the region of frequencies where collective electron modes in a fullerene can be excited.
Due to interaction with the uniform external field, $\bf{E}(\om)$, the variation of the
electron density, $\delta \rho(\bfr,\om)$, occurs on the inner and outer surfaces of the
fullerene shell.
This variation leads to the formation of the surface plasmon, which has two normal modes
of vibration, the symmetric ({\it s}) and antisymmetric ({\it a}) ones
\cite{Lambin_Lukas_1992_PRB.46.1794,Oestling_1993_EPL.21.539,Lo_2007_JPB.40.3973,Korol_AS_2007_PRL_Comment}.
Hence, the cross section, $\sigma(\om) \propto {\rm Im} \,\alpha(\om)$, is defined as
\begin{eqnarray}
\hspace{-0.5cm}
\sigma(\om) &=& \frac{4\pi \om^2}{c} \nonumber \\ &\times&
\left[
{ N_{\rm s} \,\Gamma_{\rm s} \over \bigl(\om^2-\om_{\rm s}^2\bigr)^2+ \om^2\Gamma_{\rm s}^2}
+
{ N_{\rm a} \,\Gamma_{\rm a} \over \bigl(\om^2-\om_{\rm a}^2\bigr)^2+ \om^2\Gamma_{\rm a}^2}
\right] ,
\label{CS_plasmon}
\end{eqnarray}
where $\om$ is the photon energy, $\om_{\rm s}$ and $\om_{\rm a}$ are the resonance
frequencies of the two plasmon modes,
$\Gamma_{\rm s}$ and $\Gamma_{\rm a}$ are the corresponding widths, and
$N_{\rm s}$ and $N_{\rm a}$ are the number of delocalized electrons,
involved in each collective excitation mode.
The latter values obey the sum rule $N_{\rm s} + N_{\rm a} = N$, where $N$ stands
for a total number of delocalized electrons in the fullerene.
In the present study, we account for the both $\pi$ and $(\sigma+ \pi)$ plasmons,
which involve only $\pi$ or both $\sigma + \pi$ delocalized electrons of the system,
respectively.
Thus, the photoionization cross section is defined as
$\sigma(\om) = \sigma^{\pi}(\om) + \sigma^{\sigma + \pi}(\om)$, where the contribution
of each plasmon is governed by the two modes, as follows from Eq.~(\ref{CS_plasmon}).
The frequencies of the collective excitations are defined as
\cite{Lambin_Lukas_1992_PRB.46.1794,Oestling_1993_EPL.21.539}:
%
%
\begin{eqnarray}
(\om_{\rm s/a}^{\sigma+\pi})^2 = \om_0^2 &+&
\frac{N^{\sigma+\pi}}{2 R_2^3 (1 - \xi^3)}\left( 3 \mp \sqrt{1 + 8\xi^3} \right)  \nonumber \\
(\om_{\rm s/a}^{\pi})^2 &=&
\frac{N^{\pi}}{2 R_2^3 (1 - \xi^3)}\left( 3 \mp \sqrt{1 + 8\xi^3} \right)  \ ,
\label{PRA_om_corrected}
\end{eqnarray}
where the signs '$-$' and '$+$' correspond to the symmetric and antisymmetric modes,
respectively, and
$\xi = R_1/R_2$ is the ratio of the inner to the outer radii.
The quantities $N^{\sigma+\pi} = N_{\rm s}^{\sigma+\pi} + N_{\rm a}^{\sigma+\pi}$ and
$N^{\pi} = N_{\rm s}^{\pi} + N_{\rm a}^{\pi}$ are the number of delocalized electrons
involved in the formation of the $(\sigma+\pi)$- and $\pi$-plasmons, respectively;
their sum is equal to the total number of delocalized electrons in the fullerene, $N$.
The parameter $\om_0$ comes from the Lorentz model of insulators and accounts for bound
electrons in the case of the $(\sigma+ \pi)$ plasmon~\cite{Oestling_1993_EPL.21.539}.
This parameter defines a threshold above which the free-electron picture of the
charge density becomes fully applicable.
Below $\om_0$, some of the valence electrons are treated as bound ones and,
therefore, are not involved in the formation of the plasmon excitation.
In this representation, frequency-dependent dielectric function
is defined as $\epsilon(\om) = 1 + \om_{\rm p}^2 / (\om_0^2 - \om^2)$ with
$\om_{\rm p}^2$ being the volume plasmon frequency defined by the equilibrium distribution
of electron density; it is expressed in terms of the total number of delocalized electrons
and the volume of the fullerene shell, $\om_{\rm p}^2 = 3N/(R_2^3 - R_1^3)$.
In this case, valence electrons
are assumed to be bound to their local sites by an average, isotropic force proportional
to $\om_0^2$~\cite{Lambin_Lukas_1992_PRB.46.1794}.
Therefore, only the $\pi$ electrons are active for photon energies below the threshold $\om_0$.
Above this value, all ($\sigma + \pi$) electrons become active and should be taken into consideration.
In the present calculations, we have utilized the value of $\om_0 = 13$~eV which was suggested
in Ref.~\cite{Lambin_Lukas_1992_PRB.46.1794} to reproduce the strong absorption peak in the
spectrum of C$_{60}$ around 20~eV~\cite{Hertel_1992_PRL.68.784}.
In Ref.~\cite{Oestling_1993_EPL.21.539}, a similar value of $\om_0 = 14$~eV was proposed
as an average energy of $\sigma - \sigma^*$ transitions between the bonding and antibonding
bands of cabron nanosystems with $sp^2$-hybridization.
When considering the contribution of the $\pi$-plasmon, the parameter $\om_0$ is set to zero.

In this study, we have also extended the above described formalism to investigate
collective electron excitations in multishell fullerenes under photon impact.
The general methodology for analyzing plasmon excitations formed in buckyonions
is presented in Appendix~\ref{General}.
In this extension, the behavior of electron density on each fullerene is governed
by the coupling parameter which is introduced according to geometry of the system.
This parameter stands for the separation distance between the outer surface of the
$j$th fullerene and the inner surface of the $(j+1)$th fullerene.
In Appendix~\ref{C60-at-C240}, this methodology is applied to the case of
C$_{60}$@C$_{240}$ and its current limitations are outlined.

\section{Results}
\label{Section_Results}

\subsection{Photoionization of C$_{60}$@C$_{240}$}

\begin{figure}[htb!]
\centering
\includegraphics[width=0.46\textwidth,clip]{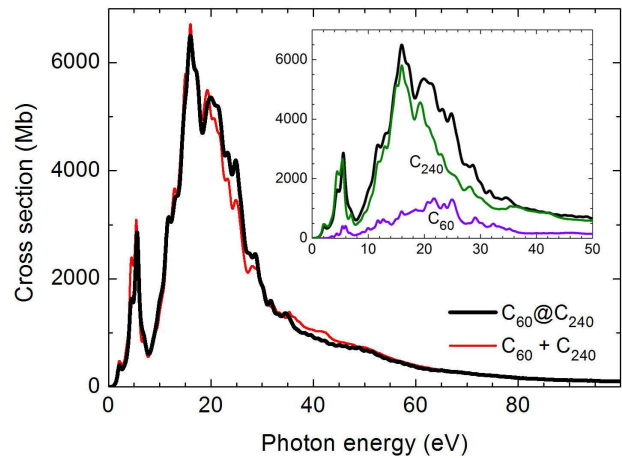}
\caption{
The photoabsorption cross section of a C$_{60}$@C$_{240}$ buckyonion calculated within
the TDDFT method (thick black curve).
Spectra of the isolated C$_{60}$ and C$_{240}$ fullerenes are shown in the inset.
The cross section $\sigma({\rm C}_{60} + {\rm C}_{240})$ which is the sum of the latter
two is shown by a thin red line.
}
\label{fig_C60@C240_TDDFT}
\end{figure}

Figure~\ref{fig_C60@C240_TDDFT} demonstrates the photoabsorption spectrum of a
C$_{60}$@C$_{240}$ buckyonion calculated within the TDDFT approach in the photon
energy region up to 100~eV (thick black curve).
The sum of the cross sections of isolated C$_{60}$ and C$_{240}$ fullerenes,
$\sigma(\om)_{{\rm C}_{60}} + \sigma(\om)_{{\rm C}_{240}}$, is shown by the thin
red curve, while the constituents of this sum are presented in the inset.
In what follows, this sum is denoted as $\sigma({\rm C}_{60} + {\rm C}_{240})$
for simplicity.
The figure demonstrates that the cross section of the buckyonion is quite similar
to that of the two isolated fullerenes.
In the photon energy range between 10 and 20~eV, the spectrum of the buckyonion
almost coincides with $\sigma({\rm C}_{60} + {\rm C}_{240})$, while a slight
difference between the two spectra appears above 20 eV.
This difference is due to a 10-15\% variation of oscillator strength and its
redistribution from the high-energy region, at about 40 eV, to the region
around 20~eV in the case of the buckyonion.
The general similarity between the calculated TDDFT-based spectrum of C$_{60}$@C$_{240}$
and the sum $\sigma({\rm C}_{60} + {\rm C}_{240})$ indicates the absence of strong plasmonic
coupling between individual fullerenes which was proposed earlier based on the jellium
model~\cite{McCune_2011_JPB.44.241002}.

The difference between our results and those reported in Ref.~\cite{McCune_2011_JPB.44.241002}
may be attributed to a different treatment of the ionic subsystem.
In the cited work, the ionic core of each fullerene comprising the buckyonion was treated as
a uniform distribution of the positive change over a spherical shell of a finite width.
On the contrary, in this work, we treat all carbon ions explicitly accounting for the icosahedral
symmetry of both C$_{60}$ and C$_{240}$, and that of the C$_{60}$@C$_{240}$ buckyonion, as well
as for their structural optimization.

\begin{figure}[htb!]
\centering
\includegraphics[width=0.46\textwidth,clip]{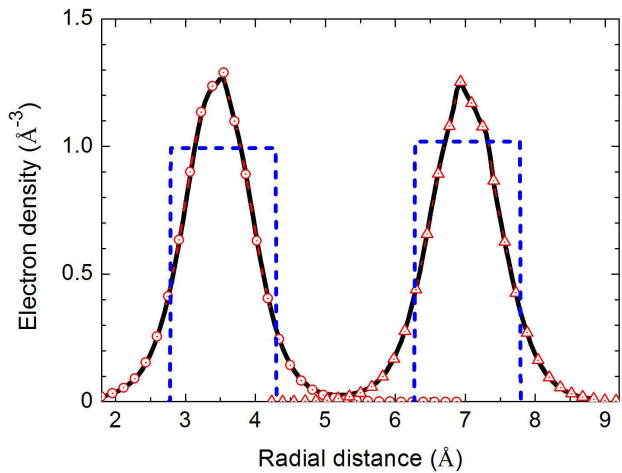}
\caption{
Radial dependence of electron density $\bar{\rho}(r)$:
density distribution of 1200 delocalized electrons in C$_{60}$@C$_{240}$ (solid black
curve) and that of 240 and 960 electrons in pristine C$_{60}$ and C$_{240}$,
respectively (symbols).
Dashed profile shows the model density distribution considered within the PRA
(see the text for further details).
}
\label{fig_C60@C240_density}
\end{figure}

To investigate the obtained results in more detail, we have analyzed the radial distribution of
valence electron density in the buckyonion and compared it to the distribution in pristine
fullerenes, see Fig.~\ref{fig_C60@C240_density}.
In the case of a fullerene C$_n$, the number $N$ of delocalized electrons represents
the four 2$s^2$2$p^2$ valence electrons from each carbon atom.
Thus, the figure illustrates the contribution of 240, 960, and 1200 electrons
in C$_{60}$, C$_{240}$, and C$_{60}$@C$_{240}$, respectively.
To calculate the density distribution, we have adopted the procedure, utilized previously
in Ref.~\cite{Verkhovtsev_2012_JPB.45.215101}.
Briefly, the electron density $\rho(\bfr)$ created by the delocalized electrons of each
system, was extracted from the Gaussian output .chk file with the help of the Multiwfn
software, ver.~3.3.8~\cite{Multiwfn}.
The density included only delocalized electrons, while the inner electron orbitals
(1$s^2$ electrons from each atom) were excluded from consideration.
Then, the electron density was averaged over the directions of the position vector~$\bfr$:
\begin{equation}
\bar{\rho}(r) = \frac{1}{4\pi} \int \rho({\bf r}) \, {\rm d}\Omega \ .
\end{equation}
Figure~\ref{fig_C60@C240_density} demonstrates that the valence electron density in the
buckyonion (solid black curve) almost coincides with that of the two pristine fullerenes (symbols).
As follows from the performed DFT calculations, the electronic structure of the buckyonion
remains almost unperturbed, compared to the isolated systems.
The binding energies of the highest-occupied molecular orbital (HOMO) and the lowest
valence MO are 5.3/24.3~eV in C$_{60}$@C$_{240}$, as compared with 5.7/24.3~eV in C$_{60}$ and
5.3/24.0~eV in C$_{240}$.
Thus, encapsulation of C$_{60}$ into C$_{240}$ does not affect the electronic properties
of the latter one and its response to the external field.
This is in agreement with the results of recent DFT-based calculations, which revealed
that the dipole polarizability of C$_{60}$@C$_{240}$ is only about 1.8\% higher than that
of the isolated C$_{240}$~\cite{Zope_2015_JChemPhys.143.084306}.
The spatial separation of the electron density on each fullerene in the buckyonion
results in a minor alteration of the photoabsorption cross section of C$_{60}$@C$_{240}$
as compared to the sum $\sigma({\rm C}_{60} + {\rm C}_{240})$.

Similar to the case of pristine C$_{240}$, the TDDFT-based spectrum of the buckyonion
is characterized by a prominent resonance peak centered at about 18~eV, which is actually
split into two narrower peaks having maximum values at 16 and 20.5~eV
(see the inset of Fig.~\ref{fig_C60@C240_TDDFT}).
This feature can be explained by the electronic structure of the larger fullerene,
namely by a dense distribution of single-electron energy levels having ionization
potential of about 10-15~eV, as follows from the performed DFT calculations.
In Ref.~\cite{Verkhovtsev_2013_PRA.88.043201}, the well-resolved features in the
photoabsorption spectrum of C$_{60}$ were assigned to discrete transitions between
particular molecular orbitals (MOs) of the fullerene, reflecting its high symmetry,
and to the ionization of particular molecular orbitals of the system.
A similar explanation should hold for the description of the two-peak profile
in the spectrum of C$_{240}$; however, a detailed analysis of the electronic
structure of this system goes beyond the scope of the present paper.
Both C$_{60}$ and C$_{240}$ have icosahedral symmetry so that their MOs are classified
according to the $I_h$ irreducible representations.
The MOs are singly ($a_g$, $a_u$), triply ($t_{1g}$, $t_{1u}$), ($t_{2g}$, $t_{2u}$),
fourfold ($g_g$, $g_u$), and fivefold ($h_g$, $h_u$) degenerated with
the subscripts ``g'' and ``u'' denoting, respectively, symmetric (``gerade'') and
antisymmetric (``ungerade'') MOs with respect to the center of inversion of the molecule.

\begin{figure}
\centering
\includegraphics[width=0.46\textwidth,clip]{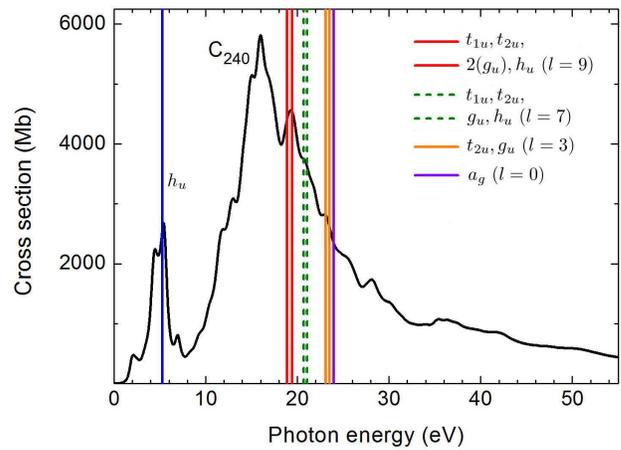}
\caption{
The photoabsorption cross section of C$_{240}$ calculated by means of TDDFT (black curve).
Vertical lines denote ionization thresholds of the HOMO, $h_u$, as well as of a number
of innermost valence MOs of the fullerene, as calculated at 6-31G(d)/LDA level of theory.
The ionization thresholds for the HOMO ($h_u$) and the innermost valence ($a_g$)
MOs are equal to 5.3 and 24.0~eV, respectively.
}
\label{fig_C240_thresholds}
\end{figure}

Due to the quasispherical structure of the C$_{60}$ and C$_{240}$ molecules,
their MOs can be expanded in terms of spherical harmonics in the angular momentum $l$
\cite{Verkhovtsev_2013_PRA.88.043201,Saito_1991_PhysRevLett.66.2637}.
For instance, the innermost valence $a_g$, $t_{1u}$, and $h_g$ MOs in the $I_h$ symmetry
represent, respectively, the $s$, $p$, and $d$ orbitals, which correspond to $l = 0, 1$, and 2.
The orbitals which correspond to higher angular momenta are constructed as a combination
of several MOs (see, e.g., Table 2 in Ref.~\cite{Verkhovtsev_2013_PRA.88.043201}).
In Fig.~\ref{fig_C240_thresholds}, we present ionization thresholds
of several particular orbitals of C$_{240}$ which are depicted by vertical lines.
The electronic structure of C$_{240}$, although being much more dense, is quite similar
to that of C$_{60}$.
For the latter molecule, we demonstrated previously~\cite{Verkhovtsev_2013_PRA.88.043201}
that there are no discrete optical transitions with the energy above 20~eV,
so that a series of peaks and bumps, arising between 20 and 25 eV, can be assigned to
the ionization of the innermost valence MOs.
As noted above, the DFT calculations for C$_{240}$, performed at 6-31G(d)/LDA level of theory,
yield the ionization potentials of the HOMO, $h_u$,
and the innermost valence MO, $a_g$, equal to 5.3 and 24.0 eV, respectively.
Thus, the features of the spectrum of C$_{240}$ in this photon energy range can be attributed
to optically allowed discrete transitions (resulting in the change of the MO's symmetry,
$g \leftrightarrow u$) and to the ionization of particular MOs of the system.

\begin{figure}
\centering
\includegraphics[width=0.46\textwidth,clip]{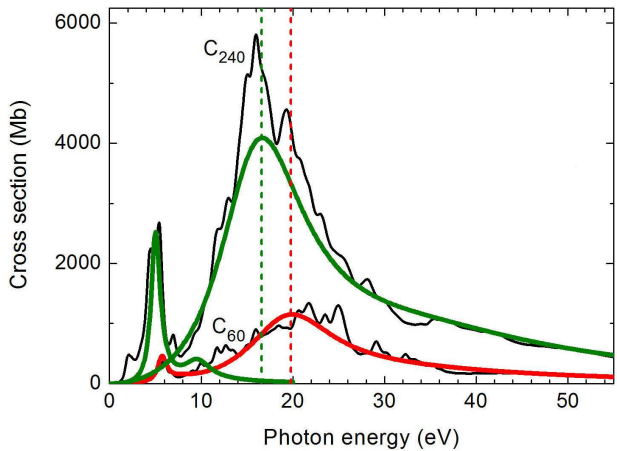}
\caption{
Contribution of the plasmon excitations to the photoabsorption cross section of C$_{60}$
(red curve) and C$_{240}$ (green curves) fullerenes, calculated by means of the PRA.
The curves, obtained within the classical approach, describe the dominating plasmon
resonance, which is formed due to collective oscillations of $\sigma$ and $\pi$ delocalized
electrons of the systems, and a narrow low-energy peak below 10 eV is attributed to
the collective excitation of only $\pi$ electrons.
Dashed vertical lines indicate the plasmon resonance frequencies $\om_{\rm s}$ in the
case of the two isolated fullerenes (see Table~\ref{C60-at-C240:Table1}).
}
\label{fig_C60_C240_PRA}
\end{figure}

\subsection{Plasmons in C$_{60}$ and C$_{240}$}

Figure~\ref{fig_C60_C240_PRA} shows the contribution of plasmon excitations to the
cross section of isolated C$_{60}$ and C$_{240}$, evaluated by means of the formalism
presented in Section~\ref{PRA_pristine_fullerene}.
The utilized parameters of the model are summarized in Table~\ref{C60-at-C240:Table1}.
In the performed analysis, we assumed that the ratio $\gamma = \Gamma/\om$ of
the width of the ($\sigma + \pi$)-plasmon resonance to its frequency is equal
to $\gamma_s = 0.6$ for the symmetric mode, and to $\gamma_a = 1.0$
for the antisymmetric mode~\cite{Bolognesi_2012_EPJD.66.254}.
These values have been utilized earlier to describe experimental data on
photoionization~\cite{Kafle_2008_JPhysSocJpn.77.014302} and electron inelastic
scattering~\cite{Bolognesi_2012_EPJD.66.254,Verkhovtsev_2012_JPB.45.141002}
of gas-phase C$_{60}$.
The value $\gamma_s = 0.6$ is also close to the numbers obtained from the earlier
photoionization and electron energy loss experiments on neutral
C$_{60}$~\cite{Hertel_1992_PRL.68.784,Mikoushkin_1998_PRL.81.2707}.
The value $\gamma_a = 1.0$ is consistent with the widths of the second plasmon
resonance observed in the photoionization of C$_{60}^{q+}$ ($q = 1 - 3$)
ions~\cite{Scully_2005_PRL.94.065503}.
Since there is no information available in the literature about the plasmon
resonance widths in C$_{240}$, we have utilized the same ratios, $\gamma_s = 0.6$
and $\gamma_a = 1.0$, as in the case of C$_{60}$.

\begin{table}[h]
\centering
\caption{Properties of the C$_{60}$ and C$_{240}$ fullerenes:
the total number of valence electrons, $N$,
the mean radius, $R$,
the inner/outer radii, $R_{1,2} = R \pm \Delta R/2$ for the thickness
$\Delta R = 1.5$~\AA,
the ratio $\xi = R_1/R_2$,
the surface plasmon energies $\om_{\rm s}$ and $\om_{\rm a}$ calculated
from Eq.~(\ref{PRA_om_corrected}),
and the corresponding widths $\Gamma_{\rm s}$ and $\Gamma_{\rm a}$,
calculated for the high-energy $(\sigma+\pi)$-plasmon and for the low-energy
$\pi$-plasmon.
Widths for the $\pi$-plasmon are taken from Ref.~\cite{Verkhovtsev_2013_PRA.88.043201}.
}
\begin{tabular}{lll}
\hline
Fullerene                  & C$_{60}$ & C$_{240}$ \\
\hline
$N$                        &   240    &     960   \\
$R$~(\AA)                  &   3.54   &     7.07  \\
$R_1$~(\AA)                &   2.75   &     6.32  \\
$R_2$~(\AA)                &   4.25   &     7.82  \\
$\xi$                      &   0.65   &     0.81  \\
%
$\om_{\rm s}^{(\sigma+\pi)}$    (eV)      &   19.8   &   16.6    \\
$\om_{\rm a}^{(\sigma+\pi)}$    (eV)      &   34.6   &   35.7    \\
$\Gamma_{\rm s}^{(\sigma+\pi)}$ (eV)      &   11.9   &   10.0    \\
$\Gamma_{\rm a}^{(\sigma+\pi)}$ (eV)      &   34.6   &   35.7    \\
$\om_{\rm s}^{\pi}$    (eV)      &   5.7   &   5.0    \\
$\om_{\rm a}^{\pi}$    (eV)      &   8.0   &   9.6    \\
$\Gamma_{\rm s}^{\pi}$ (eV)      &   \multicolumn{2}{c}{1.2~\cite{Verkhovtsev_2013_PRA.88.043201}}    \\
$\Gamma_{\rm a}^{\pi}$ (eV)      &   \multicolumn{2}{c}{3.5~\cite{Verkhovtsev_2013_PRA.88.043201}}    \\
\hline
\end{tabular}
\label{C60-at-C240:Table1}
\end{table}

Figure~\ref{fig_C60_C240_PRA} demonstrates that the PRA quantitatively describes
the main features of the spectra reasonably well.
The main resonant structure in the spectrum of each fulle\-rene is formed due to
the collective excitation of both $\sigma$ and $\pi$ electrons, while a prominent
peak in the low-energy region of the spectrum (below 10~eV) is attributed to the
excitation of some fraction of $\pi$ electrons.
Analysis of the plasmon contribution revealed that about 9 and 60 $\pi$-electrons
(out of 60 and 240 in C$_{60}$ and C$_{240}$, respectively) are involved in the
low-energy collective excitation.
The maximum of the $(\sigma+\pi)$-plasmon resonance peak for the larger fullerene is
about 3.2~eV lower than that in C$_{60}$ because of a larger size of the molecule.

The oscillator strengths for C$_{240}$, calculated by means of TDDFT and within the PRA
in the photon energy range up to 100~eV, are equal to 895 and 854, respectively.
The level of accuracy of the present calculations is similar to the earlier
calculations done for C$_{60}$~\cite{Verkhovtsev_2013_PRA.88.043201}, where the oscillator
strengths in the region up to 100~eV were estimated as 224 and 195, respectively.
In the cited paper, this difference was attributed to the contribution from single-particle
excitations, which are neglected in the model.
For the photon energies above 100~eV, the remaining oscillator strengths are
due to ionization of individual carbon atoms, multiplied by 240 and 60, respectively.
This contribution can be evaluated utilizing the asymptotic dependence of the dipole polarizability
in the region of large photon frequencies, $\alpha(\om) \propto - 1/\om^2$ \cite{Korol_AVS_BrS_2014}.

\subsection{Photoionization of C$_{20}$@C$_{60}$}

As analyzed above, the electronic and geometrical properties of C$_{60}$@C$_{240}$
do not change much as compared to the corresponding constituents.
However, it is not the case for C$_{20}$@C$_{60}$, the smallest and one of the simplest possible
buckyonions, which we have also analyzed in this work.
%
The stability as well as geometrical and electronic properties of this system
were studied earlier by means of semi-empirical, Hartree-Fock and DFT
calculations~\cite{Tuerker_2001_Theochem.545.207,Liu_2005_Theochem.725.17}.
The composite system was found to be a highly endothermic but stable structure
possessing lower symmetry compared to its isolated constituents.

\begin{figure}[htb!]
\centering
\includegraphics[width=0.46\textwidth,clip]{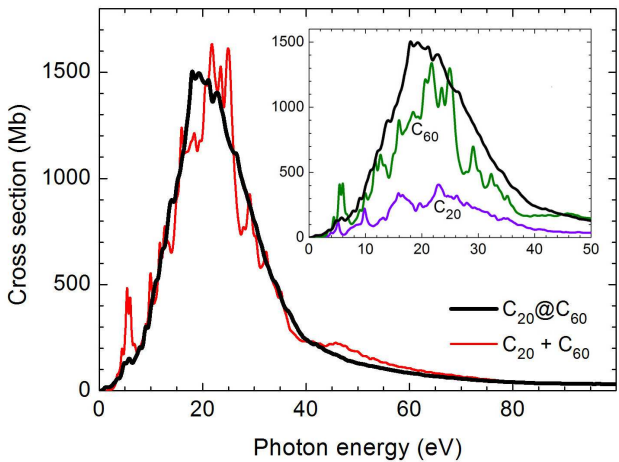}
\caption{
The photoabsorption cross section of a C$_{20}$@C$_{60}$ buckyonion calculated within
the TDDFT method (thick black line).
Spectra of isolated fullerenes C$_{20}$ and C$_{60}$ are shown in the inset.
The cross section $\sigma({\rm C}_{20} + {\rm C}_{60})$ which is a sum of the latter
two is shown by a thin red line.
}
\label{fig_C20@C60_TDDFT}
\end{figure}

The TDDFT-based cross section for C$_{20}$@C$_{60}$ is presented in
Figure~\ref{fig_C20@C60_TDDFT} (thick black curve).
Contrary to the case of C$_{60}$@C$_{240}$, the resulting spectrum of the smaller
buckyonion differs significantly from the corresponding sum of the cross sections of
isolated C$_{60}$ and C$_{20}$ (thin red curve).
The reason for this difference is that the core of the outer fullerene becomes strongly
distorted upon encapsulation of the smaller mole\-cule.
In the optimized configuration for C$_{20}$@C$_{60}$, 20 atoms of C$_{60}$ are located
in the vicinity of their initial positions and form covalent bonds with the atoms of the
smaller fullerene, while the other atoms of C$_{60}$ are pushed away from their
equilibrium position in the isolated molecule by about 0.4~\AA.
This happens because of a small inter-layer separation ($R_{{\rm C}_{20}} = 2.04$~\AA~and
$R_{{\rm C}_{60}} = 3.54$~\AA) that is comparable with the length of a ${\rm C}-{\rm C}$
single bond in the systems with $sp^2$-hybridization, $R_{\rm C-C} = 1.47$~\AA~\cite{Fox_Whitesell_OrgChem}.
Because of the fact that in C$_{20}$@C$_{60}$ the atoms of C$_{20}$ form covalent bonds
with the atoms of the C$_{60}$ cage, this structure has been suggested to be considered as
a carbon cluster rather than a multishell fullerene~\cite{Schwerdtfeger_2015_WIREs.5.96}.

\begin{figure}[htb!]
\centering
\includegraphics[width=0.46\textwidth,clip]{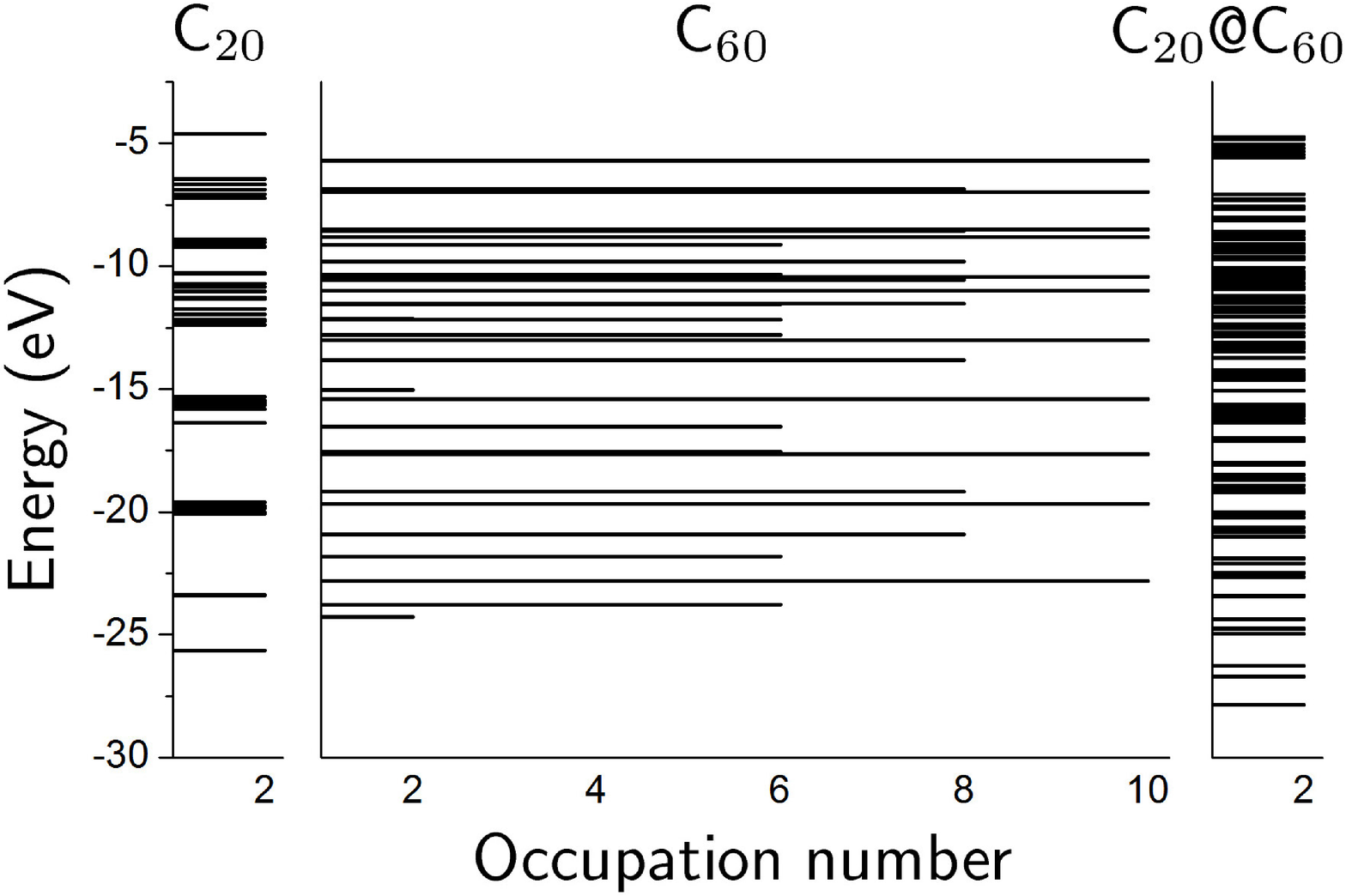}
\caption{
The ground-state electronic structure of C$_{20}$, C$_{60}$ and of C$_{20}$@C$_{60}$ buckyonion, obtained
within the quantum-mechanical framework accounting for the real symmetry of the systems.
}
\label{fig_C20@C60_elstruct}
\end{figure}

\begin{figure}[htb!]
\centering
\includegraphics[width=0.46\textwidth,clip]{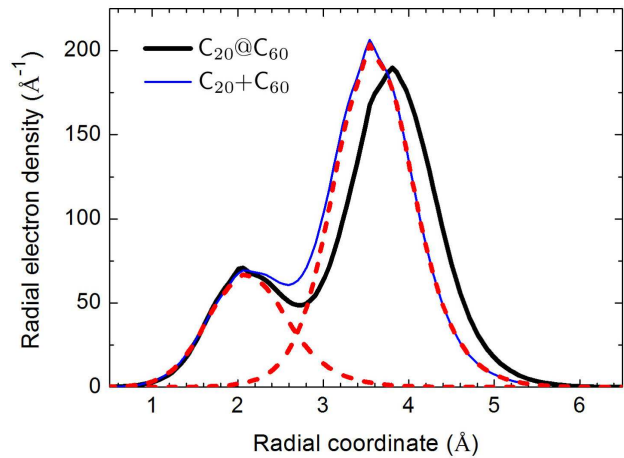}
\caption{
Radial distribution of the radial electron density, $4 \pi r^2 \bar{\rho}(r)$,
associated with the valence electrons in the C$_{20}$@C$_{60}$ (solid black
curve) and in the pristine C$_{20}$ and C$_{60}$ molecules (dashed curves).
The sum of the latter two is shown by a thin solid (blue) curve.
}
\label{fig_C20@C60_density}
\end{figure}

Geometrical distortion of the buckyonion leads to a significant rearrangement of
its electronic structure, which is illustrated in Fig.~\ref{fig_C20@C60_elstruct}.
To further support this statement, we analyzed the distribution of the valence electron density,
see Fig.~\ref{fig_C20@C60_density}.
Note that, in order to emphasize the effect, we have plotted not the average electron density
$\bar{\rho}(r)$ but the radial density, $4 \pi r^2 \bar{\rho}(r)$.
The radial density distribution reflects an increase of the volume of C$_{20}$@C$_{60}$
compared to C$_{60}$.
As a consequence, the valence electron density in the buckyonion, associated with C$_{60}$,
is shifted further from the geometrical center of the system.
The alteration of the electron density results in the complete disappearance of
the $\pi$-plasmon below 10~eV and smearing out of the fine structure atop the
($\sigma+\pi$)-plasmon (see Fig.~\ref{fig_C20@C60_TDDFT}).
Distortion of the buckyonion geometry leads to a strong reduction of symmetry, so that
the fine features, associated with discrete transitions between particular MOs of C$_{60}$,
smear out into a smoother profile.

\section{Conclusion}

This work has been devoted to the investigation of photoionization of multishell fullerenes.
Time-dependent den\-sity-functional theory was utilized to calculate the photoabsorption spectra
of C$_{60}$@C$_{240}$ and C$_{20}$@C$_{60}$ in a broad photon energy range up to 100~eV.
Apart from a minor redistribution (of 10-15\%) of the oscillator strength,
the calculated spectrum of C$_{60}$@C$_{240}$ resembles the sum of spectra of the two
isolated fullerenes, thus illustrating the absence of strong plasmonic coupling between
the fullerenes.
The absence of interplay between the fullerenes was also confirmed by analyzing radial
distribution of electron density of the system.
The calculated spectrum of the C$_{20}$@C$_{60}$ buckyonion differs significantly from the sum
of the cross sections of the individual fullerenes because of strong geometrical distortion of
the system and the related redistribution of valence electron density.

The contribution of collective electron excitations arising in individual fullerenes
was evaluated by means of plasmon resonance approximation.
It was demonstrates that the main features of the spectra, related to the formation of
plasmon excitations, are well described by means of this model approach.
An extension of the PRA formalism was presented, which allows for the study of collective
electron excitations in multishell fullerenes under photon impact.
In this extension, the behavior of electron densities on each molecule of the buckyonion
is governed by the coupling parameter which is related to the geometry of the system.
This parameter stands for the separation distance between the outer surface of the
inner fullerene and the inner surface of the outer fullerene.
As a case study, we have applied this formalism to C$_{60}$@C$_{240}$.
The performed analysis revealed a shift of the plasmon resonance frequencies in
the buckyonion compared to the case of the pristine fullerenes.
On the other hand, this shift was not observed in the TDDFT-based calculations
for C$_{240}$ and C$_{60}$@C$_{240}$.
Thus, a further analysis is required to understand the reason of this discrepancy and
how the results obtained with the analytical model can be brought in line with those
of the more elaborated method.
This investigation is of significant interest because of little knowledge on the
electronic properties of multishell fullerenes.

\section*{Acknowledgements}

A.V. acknowledges the support by the FP7 Multi-ITN Pro\-ject ``ARGENT'' (grant agreement no.~608163).
A.V.K. acknowledges the support from the Alexander von Humboldt Foundation.

\section*{Author contribution statement}

A.V. performed the calculations, analyzed the results and drafted the manuscript together with A.V.K.
A.V.S. supervised the work.
All authors discussed the results and commented on the manuscript.

\vspace{0.5cm}

\appendix
\section{Model description of the plasmon excitations in a multishell fullerene}
\label{General}

The following general equation describes the dynamic variation,
$\delta\rho(r)$, of the electron density in an arbitrary spherically
symmetric system under the action of a monochromatic uniform electric
field $\bfE_0$~\cite{Verkhovtsev_2012_EPJD.66.253,Connerade_AS_PRA.66.013207}:
\begin{eqnarray}
\left(\om^2 - 4\pi\rho_0(r)\right)\delta\rho(r)
&+&
{4\pi  \over  3}\,\rho_0^{\prime}(r)
\int_0^{\infty}\d r^{\prime}\,g(r,r^{\prime})\,
\delta\rho(r^{\prime}) \nonumber \\
&=&
\rho_0^{\prime}(r)\, \sqrt{4\pi \over 3} E_0 \ .
\label{MultiFullerene.1}
\end{eqnarray}
Here $\rho_0(r)$ is the equilibrium distribution of electron density and
$\om$ is the field frequency. The function $g(r,r^{\prime})$ is given by
\begin{eqnarray}
g(r,r^{\prime}) = \Theta(r^{\prime}-r) -
2 \left({r^{\prime} \over r}\right)^3\, \Theta(r-r^{\prime})
\label{MultiFullerene.2}
\end{eqnarray}
where $\Theta(x)$ is the Heaviside step function.

A multishell fullerene of an arbitrary level of complexity can be modeled
as a set of $n$ concentric spherical layers of finite width, i.e., a set
of individual fullerenes.
Let $R_{1j}$ and $R_{2j}$ ($R_{1j} < R_{2j}$) stand, respectively, for the
inner and the outer radii of the $j$th fullerene ($j=1,2,\dots, n$) and
$\Delta R_j = R_{2j} - R_{1j}$.
The innermost fullerene is labeled with $j=1$, and the outermost one with $j=n$.

Assuming the equilibrium distributions of electrons in each fullerene to be homogeneous,
one writes the total equilibrium density $\rho_0(r)$ in the following form
\begin{equation}
\rho_0(r) =
\sum\limits_{j=1}^{n}
\rho_{0j}\,
\Theta\left(r-R_{1j}\right)\,\Theta\left(R_{2j}-r\right).
\label{MultiFullerene.7a}
\end{equation}
Here, $\rho_{0i} = N_j/ V_{j}$ with
$N_j$ standing for the number of valence electrons in the $j$th fullerene,
and $V_{j}$ for the volume of the spherical layer,
$V_{j} = {4\pi/3}\, \left(R_{2j}^3 - R_{1j}^3\right)$.
The quantity $\rho_{0j}$ defines the plasmon frequency $\om_{\rm p j}$:
\begin{equation}
\om_{\rm p j}
= \sqrt{ 4\pi \rho_{0j} } \ .
\label{MultiFullerene.8}
\end{equation}

Since no volume plasmons can be excited under the action of a homogeneous dipole
electric field~\cite{Verkhovtsev_2012_EPJD.66.253,Korol_AS_2007_PRL_Comment},
the solution of equation~(\ref{MultiFullerene.1}) can be sought in the following form:
\begin{eqnarray}
\delta\rho(r)
=
\sum_{j=1}^n
\Bigl[
\s_{1j}\, \delta(r-R_{1j})
+
\s_{2j}\, \delta(r-R_{2j})
\Bigr]
\label{VolumeAndSurface.1}
\end{eqnarray}
where
$\s_{1j}$ and $\s_{2j}$ are
the variation of the charge densities on
the inner and outer surfaces in the $j$th fullerene.

Using (\ref{MultiFullerene.7a}) and (\ref{VolumeAndSurface.1}) in (\ref{MultiFullerene.1})
and carrying out the intermediate algebra, one derives
\begin{eqnarray}
&~&
\displaystyle{ \sum\limits_{j} }
\left[
\left( \om^2\s_{1j} + \om_{\rm p j}^2 A_{1j}\right)
\delta(r-R_{1j}) \right.
\nonumber\\
&+& \left.
\left( \om^2\s_{2j} +  \om_{\rm p j}^2 A_{2j}\right)
\delta(r-R_{2j})
\right]
\nonumber\\
&=&
\displaystyle{ \sqrt{4\pi \over 3} {E\ \over 4\pi} }
\sum\limits_{j}
\om_{\rm p j}^2
\Bigl[
\delta\left(r-R_{1j}\right)
-
\delta\left(r-R_{2j}\right)
\Bigr]
\label{MultiFullereneSurface.4}
\end{eqnarray}
where
\begin{eqnarray}
A_{1j}
&=&
-
{2\over 3}
\sum_{i<j}
\Bigl(
\s_{1i}\,\alpha_{ij}^{3}
+
\s_{2i}\,b_{ij}^{3}
\Bigr)
\nonumber \\
&-&
{1\over 3}\Bigl(2\s_{1j} - \s_{2j}\Bigr)
+
{1 \over 3} \sum_{i>j}\Bigl(\s_{1i} + \s_{2i}\Bigr)
\label{MultiFullereneSurface.2a}
\\
A_{2j}
&=&
{2\over 3}
\sum_{i<j}
\Bigl(
\s_{1i}\,a_{ij}^3
+
\s_{2i}\,\beta_{ij}^3
\Bigr)
\nonumber \\
&+&
{1 \over 3}
\Bigl(
2\s_{1j}\,\xi_{j}^3 - \s_{2j}
\Bigr)
-
{1 \over 3}
\sum_{i>j}
\left(
\s_{1i} + \s_{2i}
\right)
\label{MultiFullereneSurface.2b}
\end{eqnarray}
accompanied by
\begin{eqnarray}
&~& \alpha_{ij} = {R_{1i} \over R_{1j}},
\quad
\beta_{ij} = {R_{2i} \over R_{2j}},
\nonumber \\
&~& a_{ij} = {R_{1i} \over R_{2j}},
\quad
b_{ij} = {R_{2i} \over R_{1j}} ,
\quad
\xi_{j} = {R_{1j}\over R_{2j}} \ .
\label{MultiFullereneSurface.3}
\end{eqnarray}

For each $j$, one equalizes the terms containing identical delta-functions on
the left- and right-hand sides of~(\ref{MultiFullereneSurface.4}) and obtains
the system of $2n$ equations.
For a single fullere\-ne ($n=1$), this system reduces to the equations presented and
analyzed in Ref.~\cite{Verkhovtsev_2012_EPJD.66.253}.

\subsection{System of two concentric fullerenes, C$_{N_1}$@C$_{N_2}$
\label{n_eq_2}}

For $n=2$, the system of equations~(\ref{MultiFullereneSurface.4}) can be
explicitly written in the matrix form:
\begin{eqnarray}
\mathsf{D}
\left(
\begin{array}{c}
\s_{11}\\
\s_{21}\\
\s_{12}\\
\s_{22}\\
\end{array}
\right)
=
\left(
\begin{array}{r}
-\om_{\rm p1}^2 \, f \\
 \om_{\rm p1}^2 \, f \\
-\om_{\rm p2}^2 \, f \\
 \om_{\rm p2}^2 \, f \\
\end{array}
\right)
\label{n_eq_2.54}
\end{eqnarray}
where $f = - \sqrt{4\pi/3} E_0/4\pi$,
and $\mathsf{D}$ denotes the matrix
\begin{eqnarray}
\mathsf{D}
=
\left(
\begin{array}{cc}
\mathsf{D}_{11}   & \mathsf{D}_{12} \\
\mathsf{D}_{21}   & \mathsf{D}_{22} \\
\end{array}
\right)\,.
\label{n_eq_2.55}
\end{eqnarray}
The diagonal blocks
\begin{eqnarray}
\mathsf{D}_{jj}
=
\left(
\begin{array}{cc}
 \om^2 - 2\lambda_j  & \lambda_j  \\
 2\lambda_j \xi_j^3   & \om^2-\lambda_j     \\
\end{array}
\right),
\quad
j=1,2
\label{n_eq_2.55a}
\end{eqnarray}
describe the excitations in isolated fullerenes,
whereas the blocks $\mathsf{D}_{12}$ and $\mathsf{D}_{21}$ are due to the
interaction between the fullerenes:
\begin{eqnarray}
\mathsf{D}_{12}
&=&
\left(
\begin{array}{cc}
  \lambda_1  &\lambda_1 \\
- \lambda_1  &-\lambda_1\\
\end{array}
\right) \ ,
\nonumber \\
\mathsf{D}_{21}
&=&
\left(
\begin{array}{cc}
- 2\lambda_2 \alpha_{12}^3 &-2\lambda_2  b_{12}^3    \\
  2\lambda_2   a_{12}^3    & 2\lambda_2 \beta_{12}^3  \\
\end{array}
\right) \ .
\label{n_eq_2.55c}
\end{eqnarray}
In these formulae $\lambda_j={\om_{\rm p j}^2 / 3}$.

The determinant $|\mathsf{D}|$ is equal to
\begin{eqnarray}
|\mathsf{D}|
&=&
|\mathsf{D}_{11}| |\mathsf{D}_{22}|
\\
&-&
2\lambda_1\lambda_2\left(1 - \xi_1^3\right)\left(1 - \xi_2^3\right)
(\om^2- 2\lambda_1)(\om^2 - \lambda_2)\,
b_{12}^{3} \nonumber \ ,
\label{n_eq_2.60}
\end{eqnarray}
where $b_{12}=R_{21}/R_{12}$ and $|\mathsf{D}_{jj}|$ are determinants
of the matrices $\mathsf{D}_{jj}$:
\begin{eqnarray}
|\mathsf{D}_{jj}|
=
\left(\om^2-\om_1^{(j)2}\right)
\left(\om^2-\om_2^{(j)2}\right)
\label{n_eq_2.63}
\end{eqnarray}
where
\begin{eqnarray}
\om_1^{(j)}
=
\om_{\rm p j} \sqrt{ 3 - p_j \over 6}
\,,
\quad
\om_2^{(j)} =
\om_{\rm p j} \sqrt{ 3 + p_j \over 6}
\label{n_eq_2.64}
\end{eqnarray}
with $p_{j} = \sqrt{1+8\,\xi_j^3}$.
The frequencies (\ref{n_eq_2.64}) correspond to the symmetric,
$\om_1^{(j)}\equiv \om_{\rm s}^{(j)}$, and
antisymmetric,  $\om_2^{(j)}\equiv \om_{\rm a}^{(j)}$, modes
of the surface plasmon oscillations in a pristine C$_{N_j}$ fullerene.
For C$_{N_1}$@C$_{N_2}$, the resonance frequencies $\om_k$ ($k=1,2,3,4$)
are found as the roots of the secular equation
\begin{eqnarray}
\label{seqular_equation}
&~& |\mathsf{D}_{11}||\mathsf{D}_{22}|
\\
&-&
2\lambda_1\lambda_2\left(1 - \xi_1^3\right)\left(1 - \xi_2^3\right)
(\om^2- 2\lambda_1)(\om^2 - \lambda_2)\,
b_{12}^{3}
= 0 \nonumber \ .
\end{eqnarray}

In the limit of uncoupled fullerenes, the secular equation reduces
to $|\mathsf{D}_{11}||\mathsf{D}_{22}|=0$ resulting in~(\ref{n_eq_2.64}).
Formally, this limit corresponds to $b_{12}=0$ in equation~(\ref{seqular_equation}).
Indeed, this parameter is the only one which couples the characteristics
of both fullerenes.

\section{Application to C$_{60}$@C$_{240}$ \label{C60-at-C240}}

A C$_{60}$@C$_{240}$ buckyonion can be modeled as a set
of two concentric spherical shells of the same width $\Delta R$,
see Fig.~\ref{fig_C60@C240_density}.
The inner and outer radii of the fullerenes as well as the related parameters
are summarized in Table~\ref{C60-at-C240:Table1}.
In what follows, the indices $j=1,2$ labels the C$_{60}$ and C$_{240}$ fullerenes,
respectively.

Figure~\ref{fig_C60@C240_density} and the data presented in Table~\ref{C60-at-C240:Table1}
suggest that the values of electron densities, $\rho_{0j}$, in both fullerenes
are essentially the same, yielding the discrepancy of ca.~1.5\%.
Assuming $\rho_{01}=\rho_{02}$ one equalizes the plasmon frequencies:
\begin{eqnarray}
\om_{\rm p1}^2  = \om_{\rm p2}^2 \equiv \om_{\rm p}^2
\label{C60-at-C240:eq.01}
\end{eqnarray}
with $\om_{\rm p} \approx 37.2$ eV.

This relation allows one to solve the secular equation
analytically.
Indeed, Eq. (\ref{seqular_equation}), being written in terms of
the variable $\eta = \om^2/\om_{\rm p}^2 - 1/2$, can be further reduced
to the bi-quadratic one resulting in the following set of $\om_{k}$:
\begin{eqnarray}
\begin{cases}
\displaystyle{
\frac{\om_1^2}{\om_{\rm p}^2}
=
{2\over 9}
s_1s_2
\left[
{ s_1 + s_2 \over s_1s_2}- b^3
-
\sqrt{
\Bigl(b^3 + \chi_{-}\Bigr)
\Bigl(b^3 + \chi_{+}\Bigr)
}
\right]
}
\vspace{0.15cm} \\
\displaystyle{
\frac{\om_2^2}{\om_{\rm p}^2}
=
{2\over 9}
s_1s_2
\left[
{ s_1 + s_2 \over s_1s_2}- b^3
+
\sqrt{
\Bigl(b^3 + \chi_{-}\Bigr)
\Bigl(b^3 + \chi_{+}\Bigr)
}
\right]
}
\vspace{0.15cm} \\
\om_3^2 = \om_{\rm p}^2 - \om_2^2
\vspace{0.15cm} \\
\om_4^2 = \om_{\rm p}^2 - \om_1^2
\end{cases}
\label{C60-at-C240:eq.02}
\end{eqnarray}
where $b = R_2(\rmC_{60})/R_1(\rmC_{240})\approx0.67$,
$s_j = 1 - \xi_j^3$ and
$\chi_{\pm} =
\left(\xi_1^{3/2}\pm \xi_2^{3/2}\right)^2/s_1 s_2$.
For the purpose of self-consistency, thus calculated values of $\om_k^2$ $(k = 1 \dots 4)$
should be augmented by the additional term $\om_0^2$, introduced and explained in
Section~\ref{PRA_pristine_fullerene}.

\begin{table}[h]
\centering
\caption{
Resonance frequencies calculated for $b=0$ (the limit of uncoupled fullerenes)
and for $b=0.67$ which corresponds to the model geometry of C$_{60}$@C$_{240}$.
}
\begin{tabular}{@{}lcccc}
\hline
 $b$     & $\om_{1}$ (eV) &$\om_2$ (eV) & $\om_{3}$ (eV) &$\om_{4}$ (eV)  \\
\hline
%
0    &   16.6    &   19.8 &   34.6    &  35.7      \\
0.67 &   14.5    &   20.7 &   34.1    &  36.5      \\
\hline
\end{tabular}
\label{C60-at-C240:Table2}
\end{table}

Carrying out the limit $b=0$ in Eq.~(\ref{C60-at-C240:eq.02}),
one relates $\om_k$ to the frequencies of the symmetric and antisymmetric
surface plasmon modes in pristine C$_{60}$ and C$_{240}$:
\begin{eqnarray}
&~& \left.\om_1\right|_{b=0} = \om_{\rm s}(\rmC_{240}),
\quad
\left.\om_2\right|_{b=0} = \om_{\rm s}(\rmC_{60}),
\nonumber \vspace{0.15cm} \\
&~& \left.\om_3 \right|_{b=0}= \om_{\rm a}(\rmC_{60}),
\quad  \hspace{0.11cm}
\left.\om_4 \right|_{b=0}= \om_{\rm a}(\rmC_{240})\,.
\label{C60-at-C240:eq.04}
\end{eqnarray}

\begin{figure} [h]
\centering
\includegraphics[scale=0.35,clip]{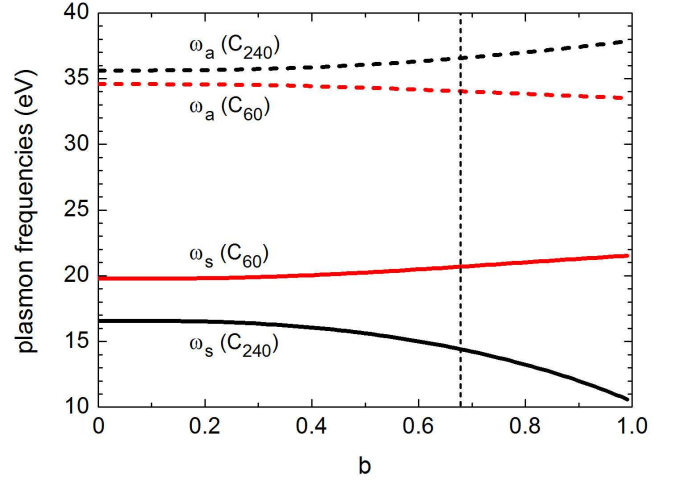}
\caption{
Resonance frequencies $\om_{1,\dots,4}$ as formal
functions of the coupling parameter $b$.
The frequencies were calculated using Eq.~(\ref{C60-at-C240:eq.02}) and
augmented by the term $\om_0^2$ as described in the main text.
The curves correspond to the plasmon frequency $\om_{\rm p}=37.2$~eV.
Dashed vertical line marks the value $b = 0.67$ which is consistent with the
chosen inner and outer radii of the fullerenes, see Table~\ref{C60-at-C240:Table1}.
In the limit $b = 0$, the frequencies $\om_{1,\dots,4}$ correspond to the indicated
frequencies of the symmetric and antisymmetric surface plasmon modes in C$_{60}$ and
C$_{240}$.
}
\label{C60-at-C240:Fig.01}
\end{figure}

The formal dependence of $\om_k$ on $b$ is presented in Fig.~\ref{C60-at-C240:Fig.01}
where the vertical line marks the value $b \approx 0.67$ consistent with the data
from Table~\ref{C60-at-C240:Table1}.
The values of $\om_k$ for this $b$ are listed in Table~\ref{C60-at-C240:Table2} where
they are compared with the resonance frequencies in pristine C$_{60}$ and C$_{240}$,
defined by Eq.~(\ref{PRA_om_corrected}).

Writing the determinant of $\mathsf{D}$ (see Eq.~(\ref{n_eq_2.55}))
as $|\mathsf{D}| = \prod_{k=1}^4 \left(\om^2 - \om_k^2\right)$,
one resolves Eq.~(\ref{n_eq_2.54}) with respect to the surface charge densities.
The result reads:
\begin{eqnarray}
&~& \hspace{-1.0cm} \s_{11}
\displaystyle{
= {x^3 - x^2 \over 3|\mathsf{D}|}\, \om_{\rm p}^8\,E_0
}
\label{Densities_02_e:eq.01} \\
&~& \hspace{-1.0cm} \s_{21}
\displaystyle{
=
-{x^3 - \lambda(3 +2s_1)x^2 +  6\lambda^2 s_1 x
 \over 3|\mathsf{D}|}\, \om_{\rm p}^8\,E_0
}
\label{Densities_02_e:eq.02} \\
&~& \hspace{-1.0cm} \s_{12}
\displaystyle{
=
{x^3 - \lambda(3  + 2 b^3 s_1)x^2  + 2\lambda^2(1 + 2 b^3) s_1 x
\over
3|\mathsf{D}|}\, \om_{\rm p}^8\,E_0
}
\label{Densities_02_e:eq.03} \\
&~& \hspace{-1.0cm} \s_{22}
=
- {1\over 3|\mathsf{D}|}
\Biggl[
x^3
+ \lambda\Bigl( 2(a^3 - \beta^3-s_2) - 3 \Bigr) x^2
\nonumber \\
&+&
2\lambda^2
\Bigl(3(\beta^3 - a^3 + s_2) + \alpha^3 - b^3 + s_1 \Bigr) x
\nonumber \\
&-&4 \lambda^3\Bigl(
\alpha^3 + \beta^3 - a^3 - b^3 + s_1 s_2
\Bigr)
\Biggr]\om_{\rm p}^8\,E_0 \ .
\label{Densities_02_e:eq.04}
\end{eqnarray}
In these formulae, $x = {\om^2 / \om_{\rm p}^2}$,
$\lambda = 1/3$, $s_j=1-\xi_j^3$, and
\begin{eqnarray}
&~& \alpha = {R_{11} \over R_{12}} = 0.43 \ ,
\quad
\beta  = {R_{21} \over R_{22}} = 0.54 \ ,
\nonumber \\
&~& a = {R_{11} \over R_{22}} = 0.35 \ ,
\quad \hspace{0.11cm}
b = {R_{21} \over R_{12}} = 0.67 \ ,
\label{Densities_01:eq.06}
\end{eqnarray}
with $R_{1j}$ and $R_{2j}$ standing for the inner and outer radii
of C$_{60}$ ($j=1$) and C$_{240}$ ($j=2$).

Once the surface densities are found, one calculates the induced dipole moment $d$:
\begin{eqnarray}
d &=& \int r^3 \Bigl[
\sigma_{11} \delta(r-R_{11})
+
\sigma_{21} \delta(r-R_{21})
\nonumber \\
&+&
\sigma_{12} \delta(r-R_{12})
+
\sigma_{22} \delta(r-R_{22})
\Bigr] \d r \ .
\label{DipoleMoment:eq.01}
\end{eqnarray}
Dividing $d$ by $E_0$, one determines the dipole polarizability $\alpha(\om)$ of the system.
The final result for $\alpha(\om)$ can be written as a sum of four resonance terms:
\begin{equation}
\alpha(\om) =
\sum\limits_{k=1}^4
\frac{\calN_k}{\om_k^2 - \om^2} \ .
\label{Polarizability:eq.02}
\end{equation}
The oscillator strengths, $\calN_{k}$, associated with the resonances $\om=\om_k$, are
\begin{equation}
\calN_k = \frac{R_{22}^3}{3} \, \om_{\rm p}^2 \, A_k \ ,
\label{Polarizability:eq.03}
\end{equation}
where
\begin{eqnarray}
&~& A_1 = \left[(\kappa_{23}-\kappa_{14})(X_1-X_4)\right]^{-1} \calA(X_1) \ ,
\nonumber \\
&~& A_2 = \left[(\kappa_{23}-\kappa_{14})(X_3-X_2)\right]^{-1} \calA(X_2) \ ,
\nonumber \\
&~& A_3 = \left[(\kappa_{23}-\kappa_{14})(X_2-X_3)\right]^{-1} \calA(X_3) \ ,
\nonumber \\
&~& A_4 = \left[(\kappa_{23}-\kappa_{14})(X_4-X_1)\right]^{-1} \calA(X_4) \ .
\label{DipoleMoment_02_A1-4:eq.01_new}
\end{eqnarray}
Here
$X_k = \om_k^2/\om_{\rm p}^2$,
$\kappa_{14} = X_1X_4$,
$\kappa_{23} = X_2X_3$,
and
\begin{eqnarray}
\calA(X_k) = a_3 X_k^3 + a_2 X_k^2 + a_1 X_k + a_0 \ ,
\label{DipoleMoment_02_A1-4:eq.06}
\end{eqnarray}
with
\begin{eqnarray}
&~& a_3 = s_2 + b^3s_1 - b^3s_1s_2 \ ,
\quad
a_2 = -{5a_3\over 3} \ ,
\nonumber \\
&~& a_1 = {2a_3 \over 3} - {3a_0 \over 2} \ ,
\quad
a_0 = - {4 \over 27}(1- b^3)s_1s_2 \ .
\label{DipoleMoment_02_A1-4:eq.07}
\end{eqnarray}

Thus defined oscillator strengths satisfy the sum rule
$\sum_{j=1}^4 \calN_j = N_1 + N_2 =1200$, that is the total number
of delocalized electrons in C$_{60}$@C$_{240}$.

In the limit of uncoupled fullerenes, the oscillator stren\-gths
$\calN_2$ and $\calN_3$ reduce, respectively, to
$N_1(p_1 + 1)/2p_1$ and $N_1(p_1 - 1)/2p_1$
(where $p_1 = \sqrt{1+8\xi_1^3}$) which stand for the number of electrons
in pristine C$_{60}$ participating in the symmetric and antisymmetric
oscillation modes \cite{Verkhovtsev_2012_EPJD.66.253,Lo_2009_PRA.79.063201}.
The quantities $\calN_1$ and $\calN_4$ reduce to those in pristine C$_{240}$.

More accurate treatment of  $\alpha(\om)$
must account for damping of the plasmon oscillations.
Formally, this can be achieved by introducing the finite widths,
$\Gamma_k$, in the denominators in (\ref{Polarizability:eq.02}):
$\om_k^2 - \om^2 \to \om_k^2 - \om^2 - \imath \om \Gamma_k$.
The widths can be calculated considering the decay of the collective excitation mode
into the incoherent sum of single-electron excitations~\cite{Gerchikov_2000_PRA.62.043201}.
With the widths introduced, the photoionization cross section of a buckyonion is found from
\begin{equation}
\sigma(\om) =
\frac{4\pi \om}{c} \, {\rm Im}\, \alpha(\om)
\label{PI_CS:eq.02}
\end{equation}
where $c$ is the speed of light and the polarizability $\alpha$ is defined
by Eq.~(\ref{Polarizability:eq.02}).


\end{document}